\documentclass[a4paper,11pt]{article}
\pdfoutput=1 % if your are submitting a pdflatex (i.e. if you have
\usepackage{jinstpub} % for details on the use of the package, please
\usepackage{siunitx}
\usepackage{lineno}
\usepackage{subcaption}
\title{SiPM array of Xenoscope, a full-scale DARWIN vertical demonstrator}

\collaboration[c]{on behalf of the
Xenoscope* team\note[*]{www.physik.uzh.ch/en/groups/baudis/Research/Xenoscope}}

\author{R.~Peres,}
\affiliation{Department of Physics, University of Zurich, Winterthurerstrasse 190, 8057 Zurich, Switzerland }

\emailAdd{rperes@physik.uzh.ch}

\abstract{The DARWIN project aims to build and operate a next-generation observatory for dark matter and neutrino physics, featuring a time projection chamber with a proposed active target of \SI{40}{t} of liquid xenon. As an R\&D facility to test fundamental components of the future detector, Xenoscope, a full-scale vertical demonstrator with $\sim$\SI{400}{kg} of liquid xenon and up to \SI{2.6}{m} electron drift length, was built at the University of Zurich. Its main objective is to demonstrate electron drift over unprecedented distances in liquid xenon --- first in a purity monitor setup with charge readout, followed by a dual-phase time projection chamber. In this second phase, an array of 48 VUV4 MMPCs from Hamamatsu (model S13371-6050CQ-02) with a 12-channel readout will be placed above the liquid xenon column and operated as a light readout for the secondary proportional scintillation signals coming from extracted electrons in the time projection chamber.
This work presents the design and development of the silicon photomultiplier array of Xenoscope, covering the structural and electronic design, sensor characterisation at cryogenic temperature and signal simulation.}


\proceeding{LIDINE 2022: Light Detection In Noble Elements\\
  21 -- 23 September, 2022\\
  Warsaw, Poland}

\begin{document}
\maketitle
\flushbottom

\section{Introduction}
\label{sec:intro}
The DARWIN~\cite{DARWIN:2016hyl} project aims to be the next-generation liquid xenon (LXe) observatory for direct dark matter detection. With a proposed active target of \SI{40}{t} of LXe and using the dual-phase Time Projection Chamber (TPC) concept for particle detection and identification, the DARWIN experiment aims to probe WIMP-nucleon cross section down to the neutrino fog~\cite{OHare:2021utq}. %At this stage, solar neutrino interactions through CE$\nu$NS and electron scattering will  have a significant impact to the overall background and in the case of the first, be indistinguishable from WIMP-nucleon interactions. 
Apart from the search for dark matter, DARWIN will also conduct other rare-events searches such as for the neutrinoless double beta decay of $^{136}$Xe, neutrino signals from core collapse supernovae and other measurements related to solar neutrinos~\cite{DARWIN:2020jme,DARWIN:2020bnc,Aalbers:2022dzr}.

In its baseline design of a cylindrical dual-phase TPC, the target medium is the liquid phase and a set of electrodes define, from bottom to top, the drift region (cathode to gate) and the extraction region (gate to anode). In the extraction region there is a liquid-gas interface, roughly equidistant from both electrodes. Upon an interaction of a particle in the LXe target, the energy transfer is split between scintillation, ionisation and heat. The prompt \SI{175}{nm} scintillation photons are detected by photosensor arrays, with the signal denoted as S1. The free electrons from ionisation are transported towards the gate due to the existing vertical electric field, $E_{\mathrm{drift}}$. Electrons reaching the gate are extracted from the liquid to the gas phase given the stronger applied field between the gate and the anode, $E_{\mathrm{extraction}}$. As a result, a process of proportional scintillation is induced, giving origin to the second, delayed, scintillation signal or S2. For a strong enough field ($\sim\SI{10}{kV/cm}$), the charge extraction efficiency from liquid to gas is almost 100\%~\cite{Xu:2019dqb}.

LXe TPCs have been leading the field of direct dark matter searches in the last decade~\cite{XENON:2018voc, LUX:2016ggv, LZ:2022ufs, XENONCollaboration:2022kmb}. However, numerous challenges are yet to be tackled to successfully scale-up the proven detector concept. The Xenoscope facility developed at the University of Zurich directly targets the challenges arising from a \SI{2.6}{m} high TPC and testing the unprecedented long electron drift length.  A full description of the facility and results from its commissioning run can be found in~\cite{Baudis:2021ipf}.

The first phase of the project was the construction, commissioning and operation of a \SI{52}{cm} high, \SI{15}{cm} diameter, purity monitor with charge readout only. LXe properties such as electron drift velocity and electron cloud longitudinal diffusion were measured and a manuscript detailing these results is under preparation. For the next stage of the project, the full-height TPC is currently being installed. In this new configuration, several sub-systems are changed or introduced: liquid level monitoring and control are added, the high-voltage (HV) supply to the cathode is redesigned to allow considerably higher voltages, and the charge readout is replaced by an array of SiPMs to detect the proportional scintillation signals produced in the extraction region of the TPC.

The following sections describe %the main features of the Xenoscope facility and TPC (section \ref{sec:xenoscope}), 
the SiPM array design, its components and characteristics (section \ref{sec:sipmarray}), the first results of the characterisation of the photosensors used in the array (section \ref{sec:charcampaign}), and the signal simulation framework developed to inform future design, operation and \mbox{physics-studies} choices (section \ref{sec:signalsim}).
\section{The top array of Xenoscope}
\label{sec:sipmarray}

\begin{figure}[b]
\centering
\begin{subfigure}{0.3\textwidth}
    \includegraphics[width=\textwidth]{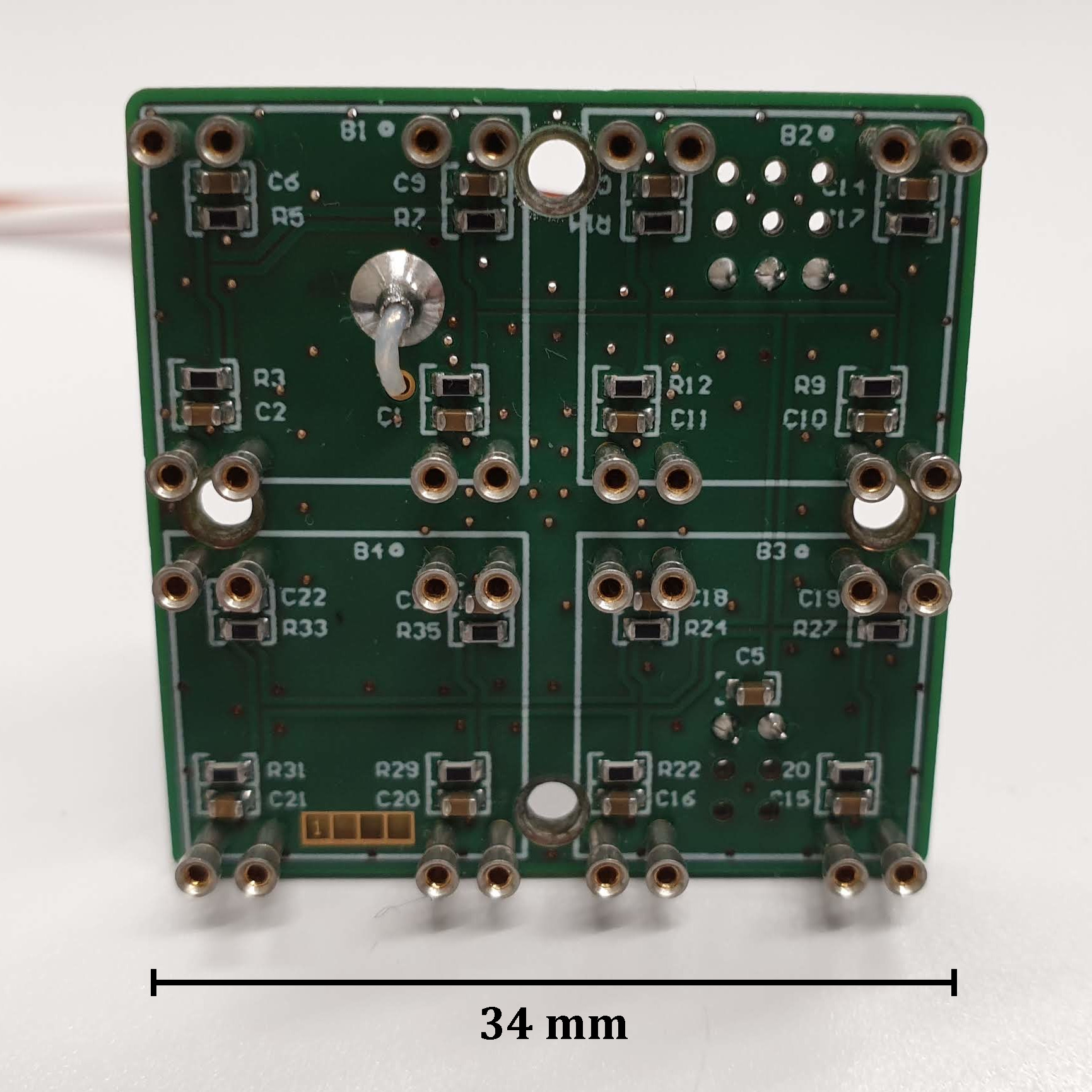}
    %\caption{}
    %\label{fig:unloadedtile}
\end{subfigure}
\hfill
\begin{subfigure}{0.3\textwidth}
    \includegraphics[width=\textwidth]{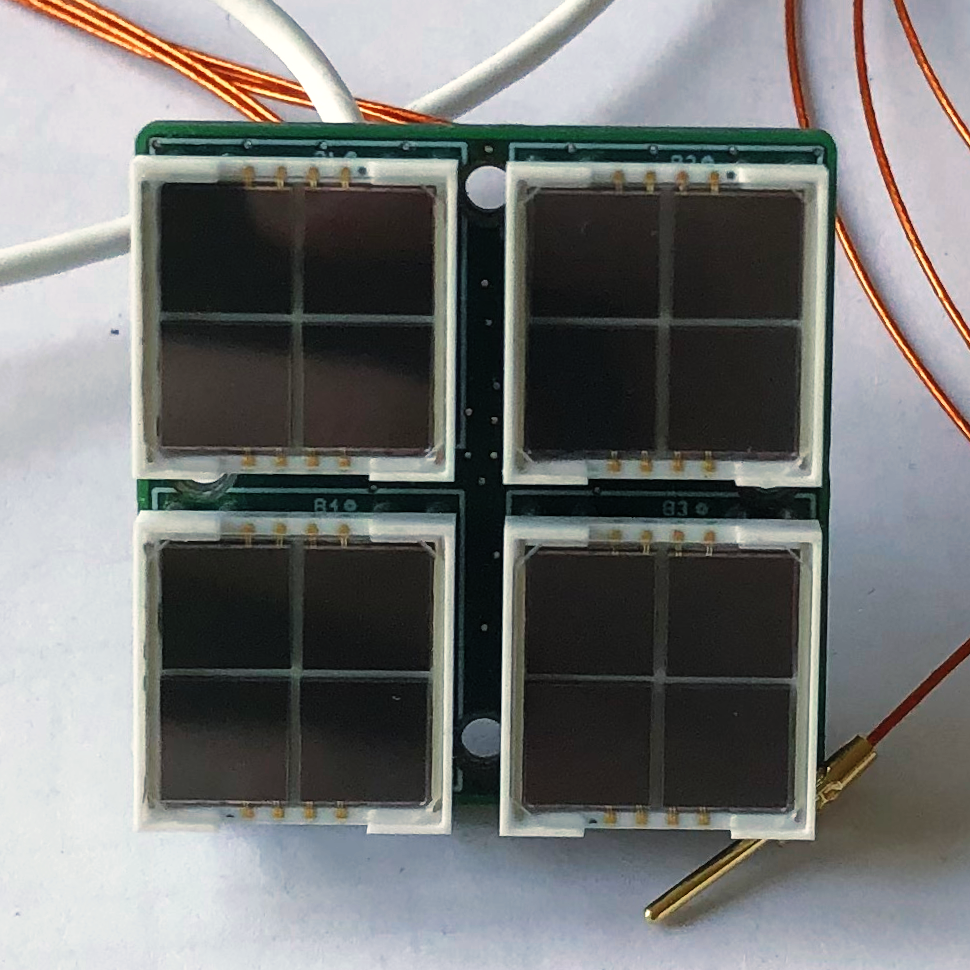}
    %\caption{}
    %\label{fig:loadedtile}
\end{subfigure}
\hfill
\begin{subfigure}{0.3\textwidth}
    \includegraphics[width=\textwidth]{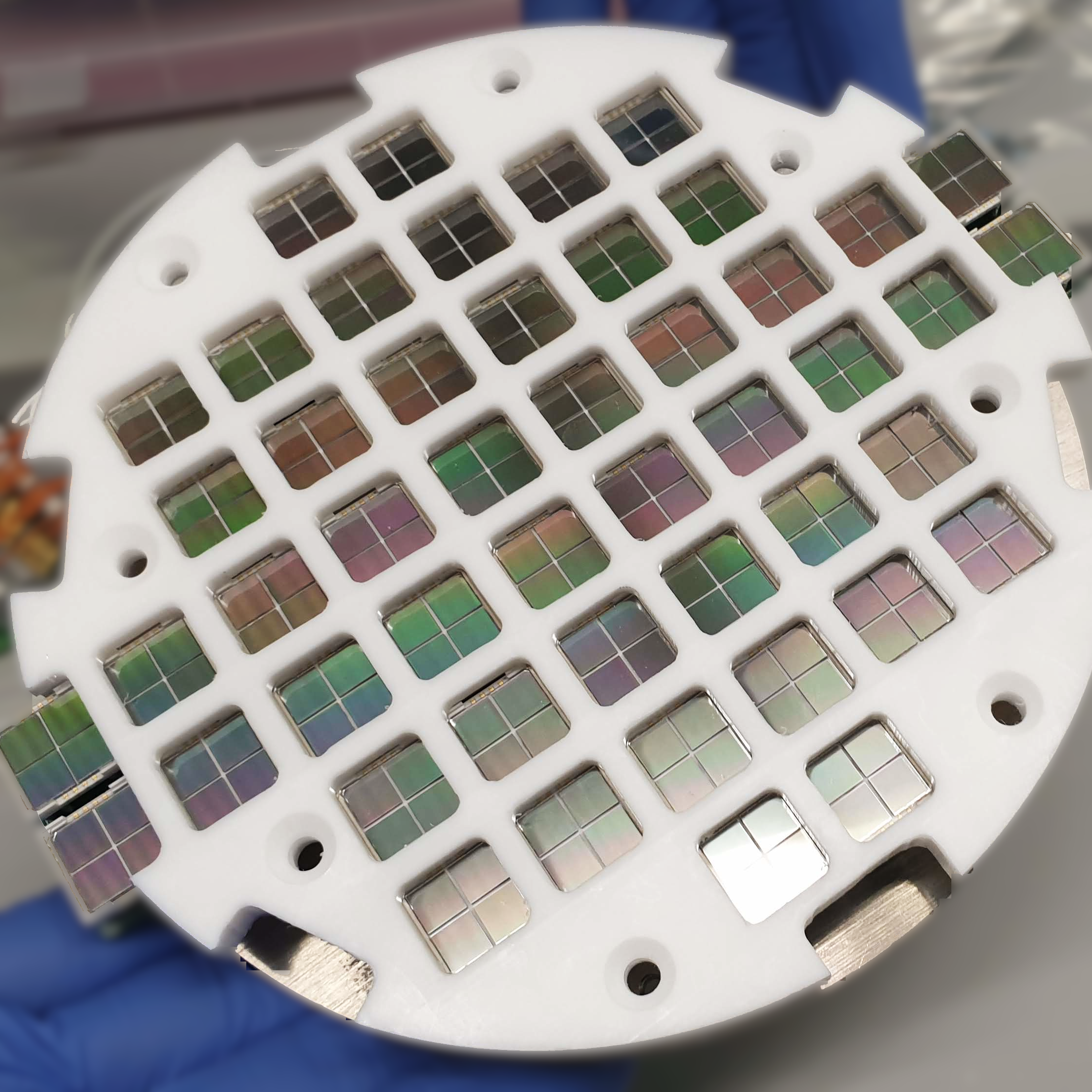}
    %\caption{}
    %\label{fig:loadedarray}
\end{subfigure}

\caption{(Left): Tile module without any MPPC modules loaded. (Centre): Tile module with four 12x12 mm$^2$ quad MPPCs. (Right) Fully loaded SiPM array with the PTFE cover.}
\label{fig:tiles}
\end{figure}

%Xenoscope will run with SiPMs in its top array as its only light sensors, focusing on recording the S2 signal of the electrons drifted from the photocathode, which are produced via photoelectric effect from the UV light of a xenon flash lamp. The 
The array consists of 192 \mbox{$6\times\SI{6}{mm^2}$} multi-pixel photon counters (MPPCs). These are packaged in $2 \times 2$ quad modules (S13371-6050CQ-02 MPPCs from Hamamatsu) with a total sensitive area of \mbox{$12\times\SI{12}{mm^2}$}. The four quads are loaded onto printed circuit boards (PCBs) with push-pin connectors, named "tiles" (figures~\ref{fig:tiles} left and centre) and There are 12 tiles in the array. In total, the array has ~$36$\% of its area covered by active sensors. The collection of tiles is screwed to a stainless steel plate for mechanical stability with Polytetrafluoroethylene (PTFE) standoffs to ensure the protection of the wiring on the backside of the PCB. When assembled in Xenoscope's TPC, the array is secured by the stainless steel plate fitting into grooves in Polyamid-imide (Torlon) pillars with the photosensors facing downward. The distance from the anode to the SiPM plane is \SI{14.65}{mm}, protecting against the risk of direct discharges through the array. To ensure that the MPPC modules are safely in place and prevent the dislodging of any units, a perforated PTFE cover of \SI{3}{mm} is placed in front of the modules (figure~\ref{fig:tiles}, right). The PTFE mask is the only reflector in Xenoscope and its goal is structural integrity. Because the primary goal is to detect O($10^3$-$10^4$) PE S2 signals from a triggered pulse, light loss due to a lack of reflectors is not a concern.

The tiles serve at the same time as holders for the SiPMs, voltage distributors and pre-amplifiers for the signals. The readout scheme is based on the design proposed in~\cite{Arneodo:2017ozr}, where the amplified output is the analogue summed signal, optimised to exclude contributions from non-triggered SiPMs. The pre-amplifier circuit is loaded with an OPA847 operational amplifier from Texas Instruments and provides a $\times$20 amplification factor to the summed signal.

\section{MPPC characterisation}
\label{sec:charcampaign}

To verify the integrity and properties of the MPPCs purchased for the array, a characterisation campaign of all the photosensor modules was conducted. The main objectives were to measure the breakdown voltage, gain, single photoelectron (SPE) resolution, dark count rate (DCR) and cross-talk probability (CTP) for each of the 48 quad sensors. %The results obtained provide the distribution of the mentioned characteristics in a large set of samples, as needed for a full scale-up of LXe TPCs photosensor arrays, and, in parallel, inform how to organise the quads in each tile in a gain matched configuration. 
This work reports the initial results from the characterisation campaign, where a fully loaded tile was used to verify the performance of the tile electronics and sensors in cold.% The complete campaign analysis and results of all 48 MPPC modules and the fully loaded array will be the subject of a future publication.

The characterisation of the photosensors was performed in the Liquid Argon Setup (LArS) at the University of Zurich using the actual array to later be installed in Xenoscope. The setup consists of a double-walled cylindrical vessel with \SI{250}{mm} diameter sealed to a top flange by an o-ring, where several feedthroughs serve the inner space with both connections to electronics, external gas bottles or vacuum pumps. The array is suspended by stainless steel rods from a PTFE holder, which, in turn, is suspended from the top flange with rods as well. In the centre of the PTFE holder, a blue LED \mbox{($\lambda = \sim~\SI{420}{nm}$)} can be used to illuminate the array. The power transmission wires of the array (kapton-insulated, stranded copper wires), which supply the bias voltage of the SiPMs and the operational amplifier, are connected to a breakout box on the inside of the vessel. This breakout box merges all the wires of the same intended purpose together. %, i.e. all the positive voltage connecting to the pre-amplifiers together, all the negative voltage connecting to the pre-amplifiers together, and so on. 
After the merge, only five wires are needed to be routed to the outside via a potted feedthrough connected on the top flange. In turn, the coax signal cables from each tile are connected directly to another potted feedthrough and are routed to the data acquisition system (DAQ).

The inner volume of the setup is first pumped-out to avoid any residual water vapour being left in the volume and then filled with helium, which is used as a coolant gas. %, which, in the particular case of this characterisation campaign, was Helium. 
The volume is then cooled via the supply of liquid nitrogen through a copper pipe coiled around the upper part of the chamber, where it undergoes liquid-to-gas transition, cooling down the system. As the pressure of the He gas decreases during cooling, more gas is supplied to keep the pressure between 1.9 and \SI{2}{bar}. The temperature inside the vessel is regulated by the flow of liquid nitrogen through the copper coil, controlled by a flow valve connected to a proportional–integral-derivative controller.

Data is acquired with two 8-channel v1724 digitizers from CAEN, read out through a v2718 VME-PCI Optical link bridge module to an on-site computer. The ADC is programmed using a custom C++ software based on CAEN libraries, while the pulse finding, data management, and processing are done with the in-house developed PyLArS package~\cite{pylars}. The software defines a signal pulse when ADC counts exceed five times the RMS of the baseline value, tuning the integration window on a per-pulse basis.

\begin{figure}[t]
\centering
\begin{subfigure}{0.315\textwidth}
    \centering
    \includegraphics[width=\textwidth]{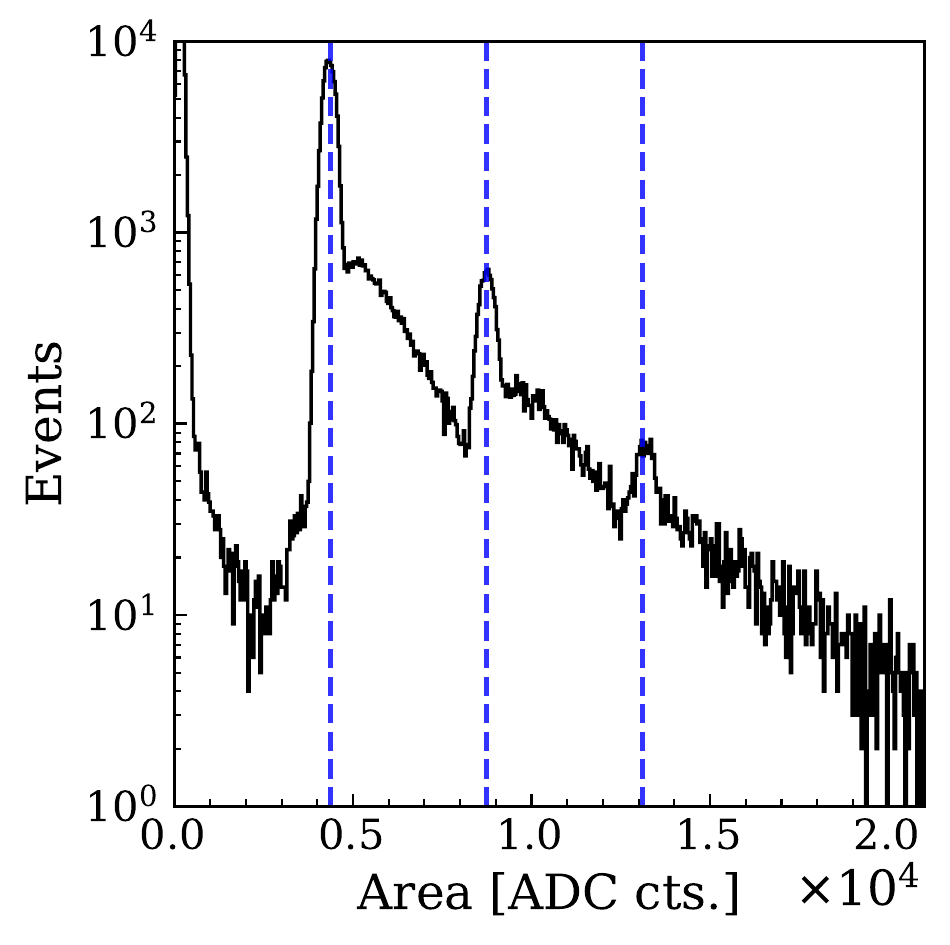}
    %\caption{}
    %\label{fig:spectrum}
\end{subfigure}
\hfill
\begin{subfigure}{0.315\textwidth}
    \centering
    \includegraphics[width=\textwidth]{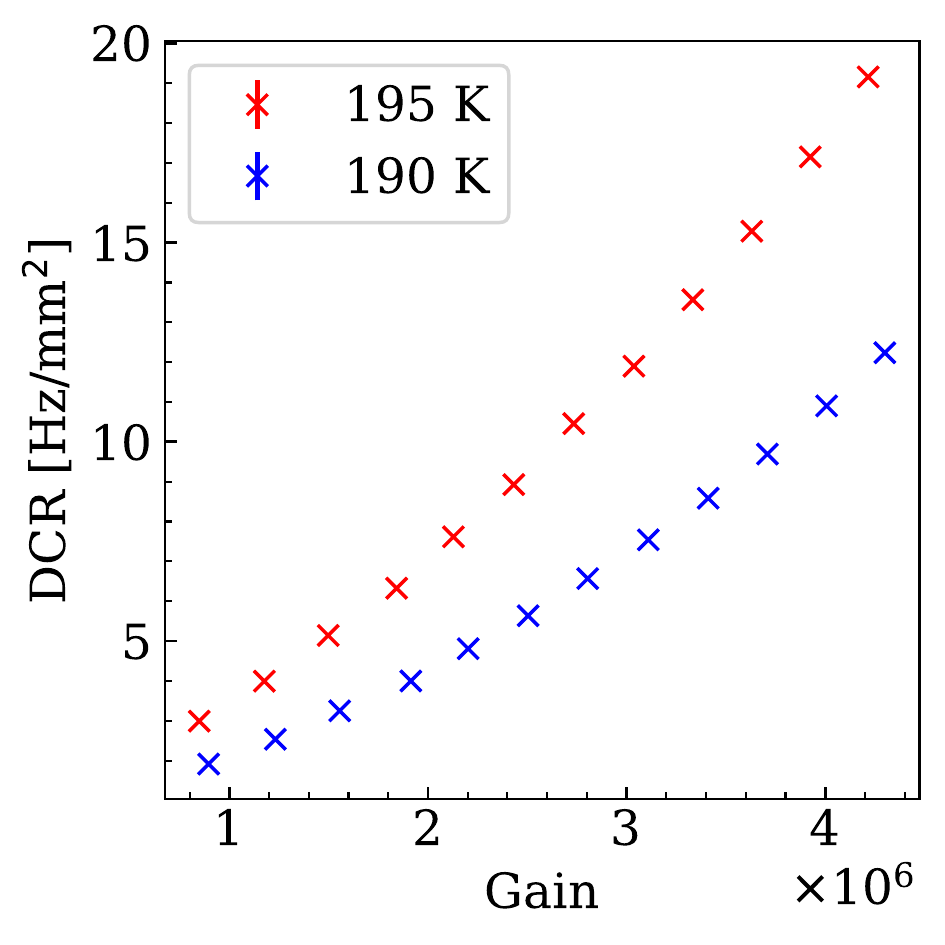}
    %\caption{}
    %\label{fig:dcr_gain}
\end{subfigure}
\hfill
\begin{subfigure}{0.315\textwidth}
    \includegraphics[width=\textwidth]{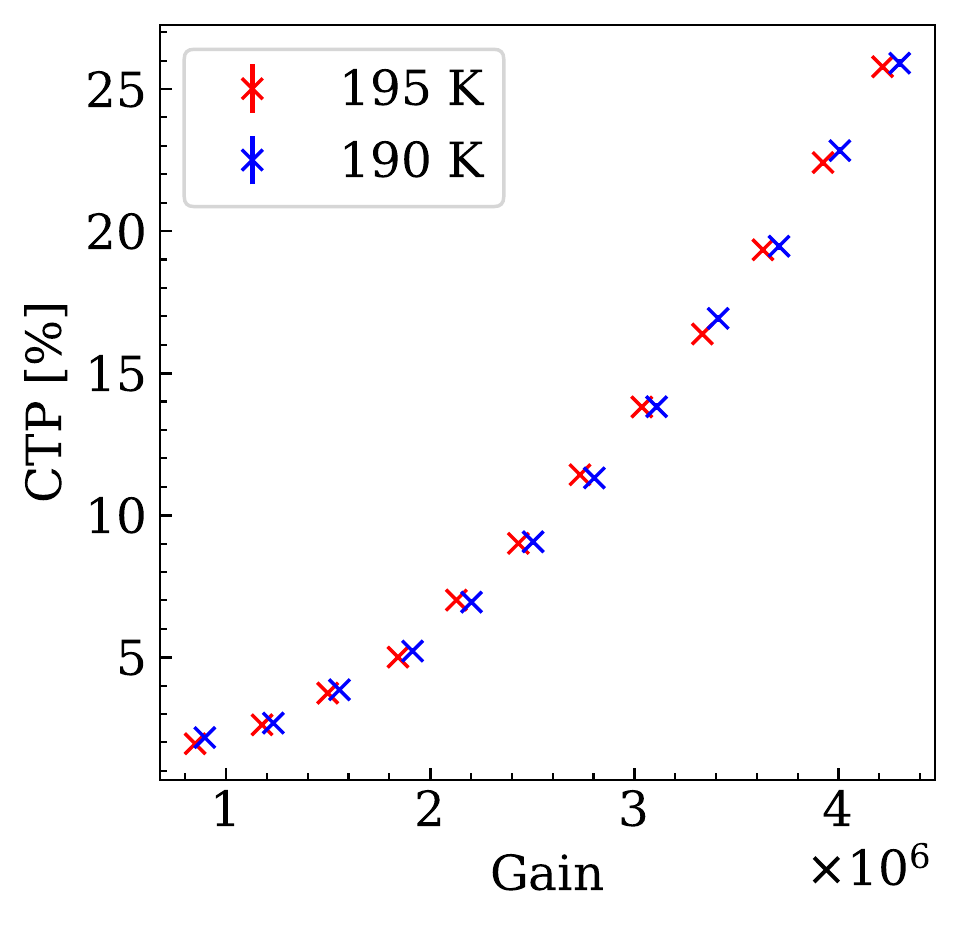}
    %\caption{}
    %\label{fig:ctp_gain}
\end{subfigure}

\caption{(Left): Area spectrum of a dark dataset at \SI{190}{K}, \SI{4}{V} over-voltage, where the 1, 2, and 3 photoelectron areas are shown in blue dashed lines. DCR (centre) and CTP (right) as a function of gain for \SI{190}{} and \SI{195}{K}.}
\label{fig:results}
\end{figure}

In figure \ref{fig:results}, left, the integrated ADC-count spectrum of a dataset in dark environment is shown after basic noise cuts on the width of the identified pulses. The first peak, result of pure dark count events, is fitted with a Gaussian function to determine the value of the SPE area and SPE resolution. Based on the SPE value, the 1, 2, and 3 PE areas are shown in blue dashed lines, aligning with the correlated cross-talk peaks. In all charge-peaks a shoulder-like region is observed on the higher-area side. The shape of such pulses mimics the shape of the main peak population, albeit with higher integrated area. These events could be due to unresolved fast correlated avalanches and were similarly observed in~\cite{Gallina:2019fxt} when studying the same family of photosensors.

At temperatures between \SI{170}{} and \SI{200}{K}, a voltage scan was performed to study its effect on the measured gain and determine the breakdown voltage. A gain of $3\times10^6$ was observed at \SI{4.5}{V} above breakdown voltage (over-voltage), corresponding to an applied voltage of \SI{52.1}{V} at \SI{170}{K} and \SI{53.2}{V} at \SI{190}{K}. As part of the gain computation, the SPE resolution is also determined. It was observed that the SPE resolution mainly depends on the over-voltage (or gain), with a small dependence on temperature. For a measured gain of $3\times10^6$, the measured SPE resolution was $\sim5\%$ at \SI{170}{K} and $\sim5.5\%$ at \SI{190}{K}.

At the temperatures for which the array is expected to be operated in Xenoscope's TPC gas-phase, i.e. between \SI{190}{K} and \SI{200}{K}, long datasets in a dark environment were taken. Defining CTP as the ratio of pulses above a 1.5 PE threshold to the pulses above 0.5 PE, the DCR and CTP were computed. The results can be found in figure~\ref{fig:results} centre and right, respectively. As expected, both quantities increase exponentially with gain. The DCR is $\mathcal{O}{(1)~\si{Hz/mm^2}}$, comparable to current~(\cite{Gallina:2019fxt}) and earlier~(\cite{Baudis:2018pdv}) versions of this  photosensor type. 

While for higher temperatures the measured DCR is higher, the CTP is observed to be independent of temperature. This is a considerable improvement from previous versions of the VUV SiPM from Hamamatsu~\cite{Baudis:2018pdv}, now $<15$\% CTP at a gain of $3\times10^6$.

\section{Signal simulation framework development}
\label{sec:signalsim}

To predict the expected signals at the SiPM array arising from ejected electrons at the photocathode with a flash of the xenon lamp, a simulation framework, XenoDiffusionScope~\cite{xenodiffusionscope}, was developed. The framework provides the base for the phenomenological study of electron longitudinal and transversal diffusion properties in the context of Xenoscope. The basic principles and steps of the simulation tool and its first results are described below.

\begin{figure}[t]
\centering

\begin{subfigure}{0.32\textwidth}
    \includegraphics[width=\textwidth]{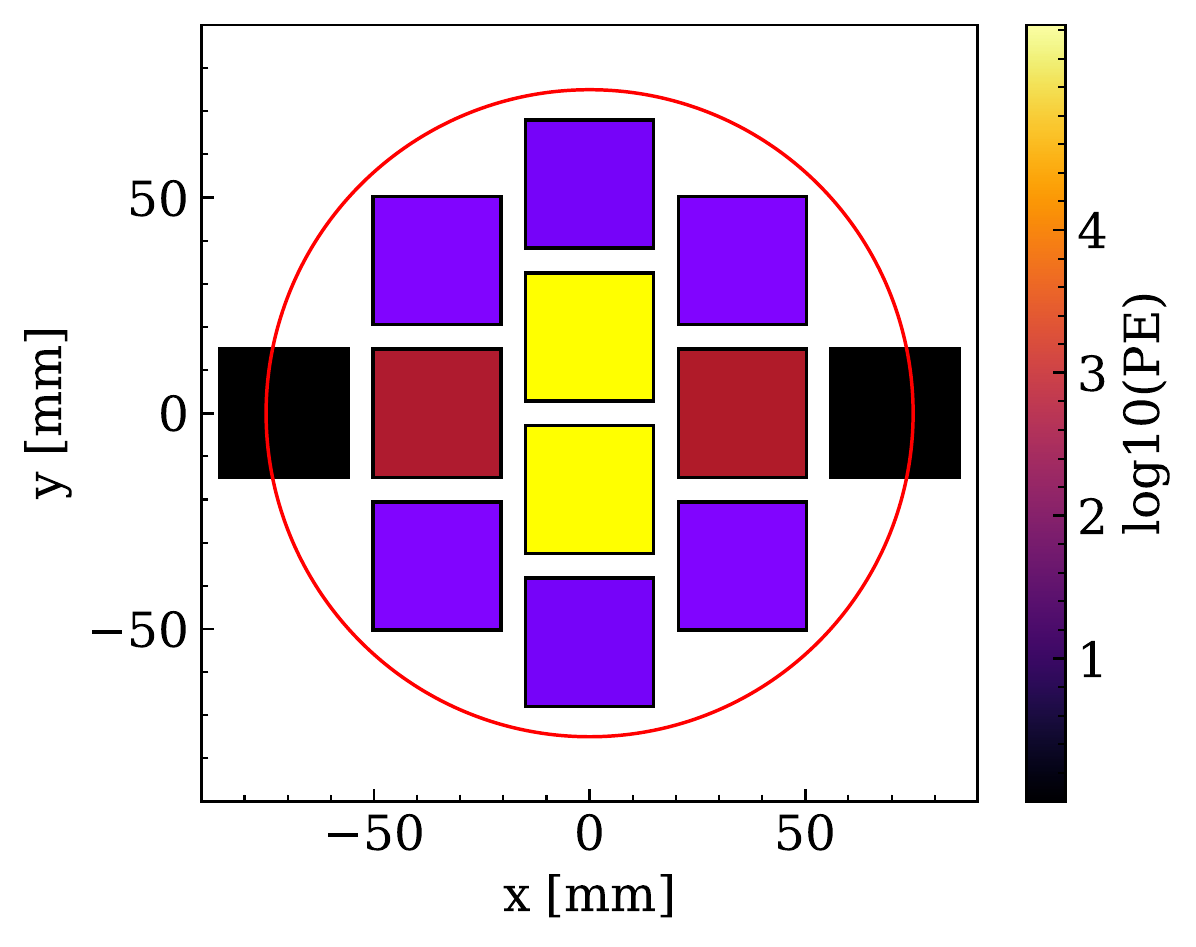}
    %\caption{}
    \label{fig:simtiles}
\end{subfigure}
\hfill
\begin{subfigure}{0.32\textwidth}
    \includegraphics[width=\textwidth]{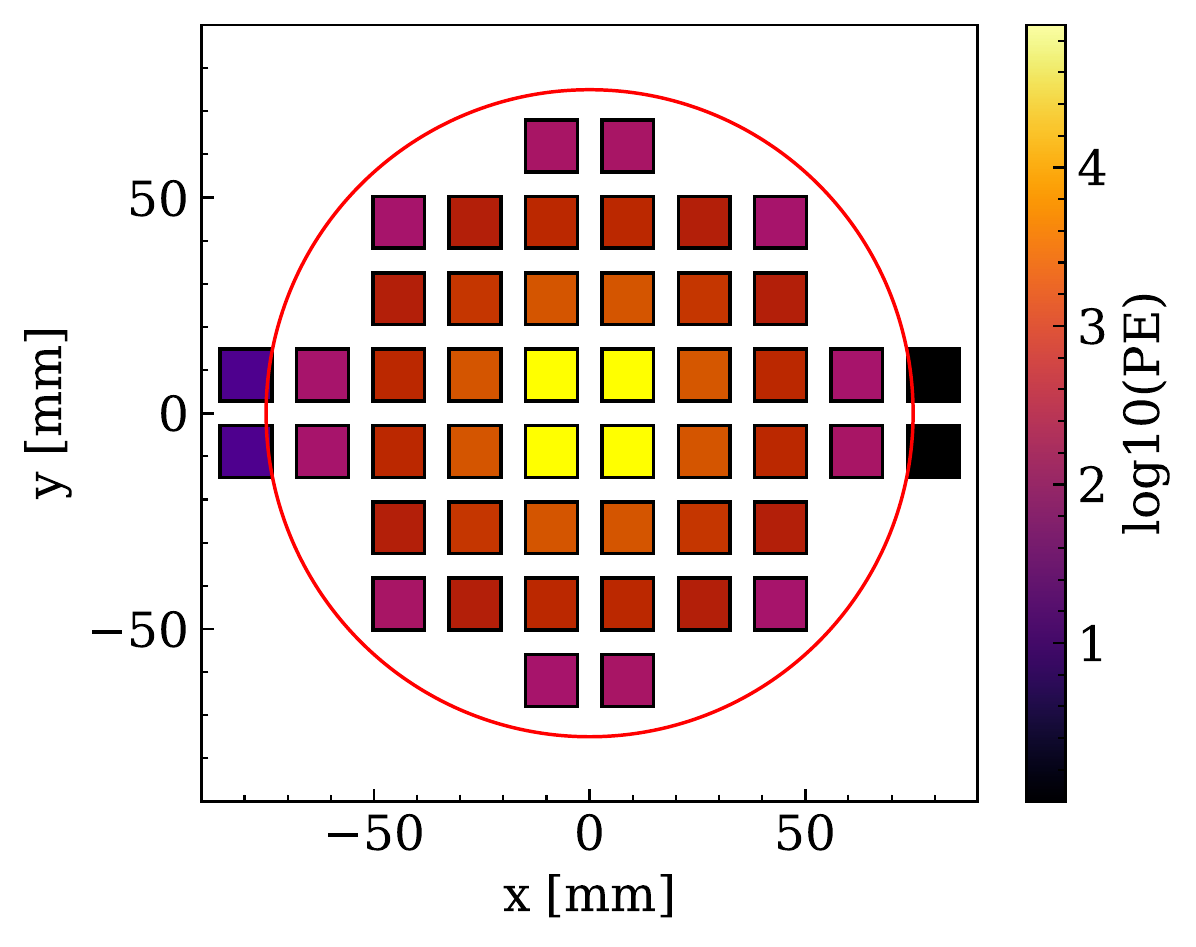}
    %\caption{}
    \label{fig:simquads}
\end{subfigure}
\hfill
\begin{subfigure}{0.32\textwidth}
    \includegraphics[width=\textwidth]{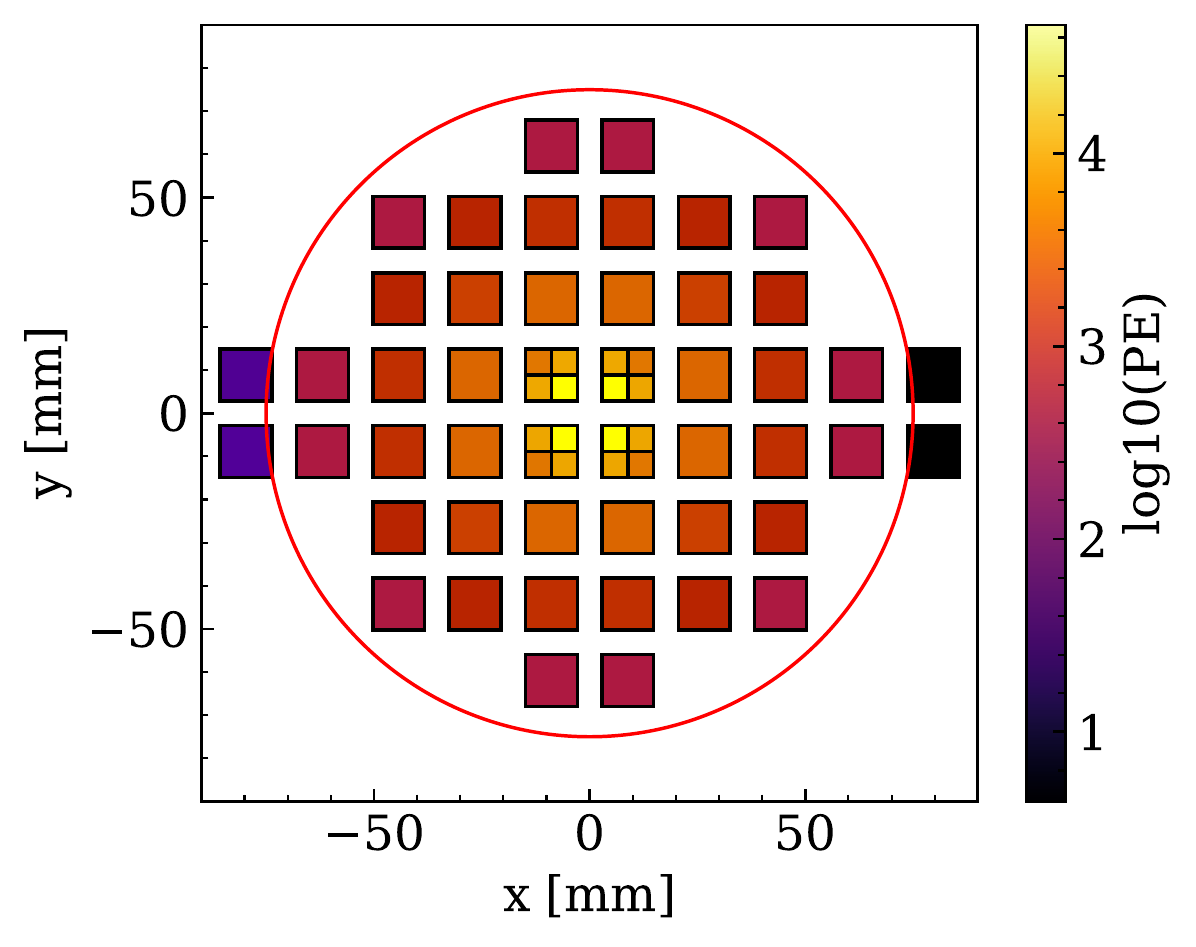}
    %\caption{}
    \label{fig:simhybrid}
\end{subfigure}

\caption{Simulated hit patterns on the top array from a flash of the xenon lamp on the photocathode for different photosensor granularity: current configuration of 12 tiles (left), individual readout channel for each quad module (centre) and an hybrid configuration with both quad modules and individual \mbox{$6\times\SI{6}{mm^2}$} units (right).}
\label{fig:simpatterns}
\end{figure}

The code is organised in a modular fashion to simplify the change of parameters and further development of new or improved sections of the toy-Monte Carlo simulator. The xenon lamp pulse is modelled to determine the initial electron cloud distribution in space and time. Once created in the LXe volume, electrons are drifted from cathode to gate, experiencing longitudinal and transversal diffusion, whose constants are defined by the user. At this stage, the boundaries of the TPC are taken into account to not propagate any electrons outside of the field cage, as well as to include electron lifetime and extraction efficiency effects. Close to the gate, the electrons are focused on the closest centre of an element of the hexagonal mesh, a known effect based on data from Xurich II~\cite{Baudis:2020nwe}. From there, the distribution of photoelectrons in the array area is computed by considering the electron-to-photoelectron gain (also known as single electron gain) and the light collection efficiency map from each of the extracted electron positions. The final pattern is achieved by summing the light patterns of all the extracted electrons and by integrating the resulting pattern in the photosensors' sensitive area.

This simulation framework will inform future design choices with respect to tiling and granularity of the SiPMs  (figure \ref{fig:simpatterns}). With constraints on correlated parameters, it can be used to study electron diffusion properties by comparing observed with simulated hit patterns. 
\section{Conclusion and outlook}
\label{outlook}

A dual-phase xenon TPC is being prepared for installation in Xenoscope, where the proportional scintillation light will be collected by an array of SiPMs in the gas phase. This work reports on the design and first results of the characterisation of the VUV4 MPPCs from Hamamatsu chosen as photosensors, their summed readout, and the developed toy-MC based signal simulation framework to inform future design choices and physics reach.

The measured gain dependence on the bias voltage is similar to the previous generation of VUV sensitive photosensors from Hamamatsu and comparable to other available photosensors. The SPE resolution was measured to be $\sim5\%$ for a gain of $3\times10^6$ in a fully loaded tile. The DCR and CTP measured are on par and improve upon the previous generation of VUV-sensitive photosensors from Hamamatsu, respectively. A shoulder-like structure was observed on the peaks of the charge spectrum, populated by events with similar shape and duration as SPE or 2 PE signals but with higher integrated charge. The results of the characterisation of the full set of 48 MPPCs and fully loaded array are in preparation and will be reported in a future publication.

\acknowledgments
This work was supported by the European Research Council (ERC) under the European Union’s Horizon 2020 research and innovation programme, grant agreement No. 742789 (Xenoscope), by the SNF Grant 200020-188716 and by the University of Zurich.

\newpage
\bibliographystyle{jhep}
\bibliography{bibliography}

\end{document}